\documentclass[aps,prd,twocolumn,showpacs]{revtex4}
\usepackage{graphicx}
\usepackage{latexsym}
\usepackage{amsmath}
\usepackage{amsfonts}
\usepackage{amssymb}
\newcommand{\be}{\begin{equation}}
\newcommand{\ee}{\end{equation}}
\newcommand{\ben}{\begin{eqnarray}}
\newcommand{\een}{\end{eqnarray}}
\begin{document}

\title{Confinement from new global defect structures}
\author{D. Bazeia$^{a}$, F.A. Brito$^{b}$, W. Freire$^{c}$
and R.F. Ribeiro$^{a}$} \affiliation{\centerline{$^a$Departamento
de F\'\i sica, Universidade Federal da Para\'\i ba,}
\\
\centerline{Caixa Postal 5008, 58051-970 Jo\~ao Pessoa, Para\'\i
ba, Brazil}
\\
\centerline{$^b$Departamento de F\'\i sica, Universidade Federal
de Campina Grande,}
\\
\centerline{58109-970 Campina Grande, Para\'\i ba, Brazil}
\\
\centerline{$^c$Departamento de Matem\'atica, Universidade
Regional do Cariri,}
\\
\centerline{63040-000 Juazeiro do Norte, Cear\'a, Brazil}}

\date{\today}

\begin{abstract}
We investigate confinement from new global defect structures in
three spatial dimensions. The global defects arise in models
described by a single real scalar field, governed by special
scalar potentials. They appear as electrically, magnetically or
dyonically charged structures. We show that they induce
confinement, when they are solutions of effective QCD-like field
theories in which the vacua are regarded as color dielectric media
with an anti-screening property. As expected, in three spatial
dimensions the monopole-like global defects generate the Coulomb
potential as part of several confining potentials.
\end{abstract}


\maketitle

\section{Introduction}

We can describe the physics of {\it heavy} quarks by using models
that engender non relativistic confining potentials. One can
successfully obtain the whole mass spectrum of the quark
anti-quark pair in the {\it quarkonium} system using, for
instance, the well-known Cornell potential \cite{cornell}
$U_C(r)\!=\!-\frac{a}{r}+br$, where $a,\,b $ are nonnegative
constants and $r$ is the distance between quarks. Although there
are many different forms of such confining potentials the most
probable potential to describe heavy quarks in the {\it
bottomonia} region, as has been shown in \cite{MZ,Z}, is
$U_{MZ}(r)\!=\!C_1(\sqrt{r}-\frac{C_2}{r})$, where the constants
are fixed as $C_1\simeq 0.71\,{\rm GeV}^{1/2}$ and $C_2\simeq
0.46\,{\rm GeV}^{3/2}$ --- see \cite{slusa} for further details.

As one knows the color vacuum in QCD has an analog in QED. In QED
the {\it screening effect} creates an effective  electric charge
that increases when the distance $r$ between a pair of
electron-antielectron decreases. On the other hand, in QCD there
exists an {\it anti-screening effect} that creates an effective
color charge which decreases when the distance $r$ between a pair
of quark-antiquark decreases. We can summarize this discussion by
writing $E(r)\sim \frac{q'(r)}{r^2}\sim\frac{q}{G(r)r^2}$, where
$E(r)$ stands for the electric (or color electric) field due to a
charge $q$ embedded in a polarizable medium characterized by a
``(color) dielectric function'' $G(r)$ and $r$ represents an
arbitrary position on the (color) dielectric medium. The QED
vacuum behavior manifests according to the effective electric
charge $q'(r\gg d)\leq1$ or $G(r\gg d)\geq1$ and $q'(r\ll d)>1$ or
$G(r\ll d)<1$, where $d$ is the typical size of the polarized
``molecules''. On the other hand, the QCD vacuum behavior
manifests according to the effective $color$ electric charge
$q'(r\gg R)\gg1$ or $G(r\gg R)\ll1$ and $q'(r< R)\to1$ or
$G(r<R)\to1$, where $R$ is the typical radius of a particular
hadron made of quarks. Since we are interested on confinement we
shall focus our attention on the latter case, where the behavior
of $G(r)$ is clearly chosen such that the QCD vacuum provides
absolute color confinement. Note that formally we can continue
using Abelian gauge fields just as in QED, but now $G(r)$ is
chosen properly in order to provide confinement. In fact, as we
shall see below, even though we consider non-Abelian gauge fields
there are some reasons to consider only the Abelian projections
when these fields are embedded in color dielectric medium
\cite{willets}. We account to these facts to study confinement of
quarks and gluons inside hadrons by using an effective
phenomenological QCD-like field theory \cite{mit,slac,cornell,fl}.

As an example consider $G(r)\!=\!\frac{a}{a+br^2}$. Notice that
$G(r\!\to\!\infty)\!\to\!0$ and $G(r\!\to\!0)\!\to\!1$ as expected
to provide perfect confinement. Using this function $G(r)$ and the
fact that $E(r)\!\sim\!\frac{q}{G(r)r^2}$, it is not difficult to
conclude that the potential $U(r)$ has the form of the Cornell
potential. As we have stated above, there are many different forms
of confining potential and then many different color dielectric
functions $G(r)$ can be used to describe confinement. However, for
simplicity we choose $G(r)$ as simpler as possible. By considering
specific color dielectric functions, we shall investigate later
models developing several possibilities of confining potentials.

In general, the behavior of the color dielectric function $G$ with
respect to $r$ can be driven by some scalar field $\phi(r)$ that
describe the dynamics of the color dielectric medium. Applying the
Lagrangian formalism to describe both dynamics of gauge and scalar
fields embedded in a color dielectric medium we have the effective
lagrangian ${\cal
L}_{eff}\!=\!-\frac{1}{4}\,G(\phi)F_{\mu\nu}F^{\mu\nu}+
\frac{1}{2}\partial_\mu\phi\partial^\mu\phi-j_\mu A^\mu$, where
$\mu,\nu\!=\!0,1,2,...,D$ and $j_\mu$ is an external color current
density. Considering static gauge fields, i.e.,
$A_\mu\!=\![U(r),0,...,0]$, $E(r)\!=\!F_{0r}$, $D\!=\!3$ and
$j_0\!=\!\rho(r)\!=\!4\pi q\delta(r)$, the equation of motion for
the gauge field gives $E\!=\!\frac{q}{G(\phi)r^2}$ where the
function $G(\phi)$ defines the confinement. The color electric
charge can be given by the fermionic sector as
$j_\mu=q\bar{\psi}\gamma_\mu\psi$ of a QCD-like theory that we
consider in the next section with further detail. Let us now make
some comment about the scalar field sector and also consider other
assumptions. Theories involving gauge and scalar fields like the
effective lagrangian ${\cal L}_{eff}$ above have been well
explored in the literature \cite{slusa,dick1,dick2,dick3}.
Specially in \cite{dick1} a model involving Abelian projections of
non-Abelian gauge field coupled to a dynamical scalar field as in
the effective Lagrangian above was considered. The scalar field
$\phi$ was identified with dilaton field whose solution of
equation of motion behaves like $\phi(r)\sim\ln{(1+\frac{a}{r})}$.
The scalar potential $V(\phi)$ is set to zero. It was shown in
this model that the Coulomb potential is regularized at short
distance. Other investigations as given in
\cite{slusa,dick2,dick3} usually consider a non-zero scalar
potential like $V(\phi)\sim \alpha\phi^{\,\beta}$ which has a
unique minimum. We consider below a different perspective of
investigation.

Let us consider the possibility of formation of defects in the
color dielectric medium. In order to implement this phenomenon the
potential $V(\phi)$ should have a set of minima whose topology is
nontrivial. These defects can be understood as nonlinear
excitations of the color dielectric medium. Since they become
charged via fermion zero modes
--- see details in Sec.~\ref{confD} --- they carry ``color''
charges localized along their spatial extensions. We investigate
the confinement of these defects since they are charged objects
embedded in a color dielectric medium.

In Classical Electrodynamics, as one knows, there are extended
charged objects that play the role of confinement. As an example,
consider the {\it infinite plane} of charges that produces the
``confining potential'' $U(r)\sim \sigma r$. Notice that this
object interacts with charged particles --- or with another
parallel infinite plane
--- in such a way that we cannot separate them from each other
since the energy of the system increases with the separation $r$.
However, other objects with {\it finite size} as a charged sphere
or radius $R$ produces an electrical field (for $r\!>\!R$) just as
the the electrical field of a point particle, i.e., $E\sim
\frac{Q}{r^2}$, where $Q$ is the total charge of the sphere. Of
course, in this case there is no confining behavior since the
energy decreases with the separation $r\!>\!R$. On the other hand,
as we have stressed above, when such charges are embedded in a
dielectric medium with anti-screening property, i.e., in the QCD
vacuum, the confinement appears.

Consider a system with gauge fields embedded in medium that
develops both dielectric and magnetic property given by the
Lagrangian ${\cal
L}\!=\!-\frac{1}{4}G(\phi)F_{\mu\nu}F^{\mu\nu}+j_\mu A^\mu
-\frac{1}{4}H(\phi)\widetilde{F}_{\mu\nu}\widetilde{F}^{\mu\nu}
+j_{m\mu}\widetilde{A}^\mu+\frac{1}{2}
\partial_\mu\phi\partial^\mu\phi$, where $j_{\mu}$ and $j_{m\mu}$
are current densities of electric and magnetic charges either
associated to point-like sources or extended charged defects.
Assuming that the dual gauge field with strength tensor
$\widetilde{F}_{\mu\nu}$ describes magnetic monopoles on the QCD
vacuum, we expect that ``magnetic permeability'' $H(\phi)$ gets
stronger as the dielectric function $G(\phi)$ gets weaker and
vice-versa --- recall that magnetic and electric charges, $h$ and
$q$, are related as $h\sim1/q$. Let us now implement the idea
proposed by 't Hooft and Mandelstam \cite{tman} --- see also
Seiberg-Witten theory \cite{sbwit} --- concerning confinement of
quarks. It considers magnetic monopoles in the QCD vacuum
generating screening currents that confine the color electric flux
in a narrow tube. This is dual  to the Abrikosov flux tube
produced in a superconductor --- see, e.g., \cite{singh} for
further discussions. In order to produce confinement,  both
$G(\phi)$ and $H(\phi)$ should have the suitable behavior  $G(r\gg
R)\to0$ ($H(r\gg R)\to1)$, for the {\it confined phase}; and
$G(r\lesssim R)\to1$ ($H(r\lesssim R)\to0)$, for {\it deconfined
phase}. $R$ is the radius of the hadron in deconfined phase and
$\phi$ is uniform in both regime, i.e., $\partial_r\phi(r)\sim0$.
In the confined phase a hadron (such as a {\it quark-anti-quark}
pair or a {\it defect-anti-defect} pair) looks like a narrow flux
tube (connecting the two sources) embedded in a monopole
condensate. This phenomenon is governed by the dual effective
Lagrangian $\widetilde{\cal
L}_{eff}\!=\!-\frac{1}{4}\widetilde{F}_{\mu\nu}\widetilde{F}^{\mu\nu}
+j_{m\mu}\widetilde{A}^\mu$, whose equations of motion are (i)
$\nabla{.}\vec{\widetilde{B}}\!=\!\rho_m$ and (ii)
$-\nabla\times\vec{\widetilde{E}}\!=\!\vec{j}_m$, where
$\vec{\widetilde{E}}$ and $\vec{\widetilde{B}}$ are static fields
and $j_\mu\!=\!0$. The homogeneous equation
$\nabla{.}\vec{\widetilde{E}}\!=\!0$ describes the uniform
electric field in the confined phase. The persistent currents in
the monopole condensate are governed by the dual London equation
$\lambda^2\nabla\times\vec{j_m}\!=\!\vec{\widetilde{E}}$, where
$\lambda$ is the London penetration depth \cite{singh}. Combining
this equation with Eq.(ii) for the electric field
$\vec{\widetilde{E}}$ we find the {\it fluxoid quantization
relation}
$\int{\vec{\widetilde{E}}.d\vec{S}}-\lambda^2\oint{\vec{j}_m.d\vec{{\ell}}}\!=\!n\Phi_e$
\cite{singh}, where $n$ is an integer and $\Phi_e\!=\!q$ is the
{\it quantum of electric flux}.

On the other hand, in deconfined phase the magnetic monopoles are
dilute and the magnetic current is negligible, i.e.,
$j_{m\mu}\!\sim\!0$. In this regime the theory is described by the
effective Lagrangian ${\cal
L}_{eff}\!=\!-\frac{1}{4}{F}_{\mu\nu}{F}^{\mu\nu}+j_\mu A^\mu$,
whose equations of motion are (iii) $\nabla{.}\vec{E}\!=\!\rho$
and (iv) $\nabla\times\vec{B}\!=\!\vec{j}$, where $\vec{E}$ and
$\vec{B}$ are static fields and the electric current $\vec{j}$
vanishes. The equation (iii) should describe the Coulomb potential
in the deconfined phase. The above equations of motion comprise
part of the full set of equations of motion we obtain from the
original Lagrangian for the scalar, electric and magnetic fields
--- fermion fields will also be included later. In deconfined
phase, since there are some magnetic monopoles, they are around
the surface of the hadron. The magnetic field due to such magnetic
monopoles can be found by using the ``Gauss law'' (i) in
$D$-spatial dimensions given as (v)
$\frac{1}{r^{D-1}}\frac{d}{dr}\left(r^{D-1}
H(\phi({r}))\widetilde{B}(r)\right)\!=\!\rho_m$. Here we assume
the magnetic charge density $\rho_m$ has radial symmetry.  We also
consider a regime $r\sim R$ in which the scalar field $\phi$ is
dynamical. In such regime, taking into account the discussions
above, and eliminating $\widetilde{B}$ through equation of motion
(v) o from the original Lagrangian gives \cite{footnote} ${\cal
L}_{eff}\!=\!-\frac{1}{4}G(\phi){F}_{\mu\nu}{F}^{\mu\nu}+j_\mu
A^\mu +\frac{1}{2}\partial_\mu\phi\partial^\mu\phi-V(r,\phi)$,
where
$V(r,\phi)\!\equiv\!V(\phi)/r^{2D-2}\!=\!Q_m^2/2\,r^{2D-2}H(\phi)$,
and $Q_m$ is the total magnetic charge of dilute magnetic
monopoles distributed on the surface of the hadron with radius
$r\!\sim\!R$ in deconfined phase. {Notice that although the
original Lagrangian ${\cal L}$ is Lorentz invariant, in deconfined
phase the scalar sector of the effective Lagrangian, i.e., ${\cal
L}^\phi_{eff}\!=\!\frac{1}{2}\partial_\mu\phi\partial^\mu\phi-V(r,\phi)$,
which supports defects with only radial symmetry, effectively
breaks Lorentz invariance. Once created, these defects get charged
electrically through fermion zero modes $\psi_0$ and then turn out
to interact with the gauge field sector ${\cal
L}_{eff}\!=\!-\frac{1}{4}G(\phi){F}_{\mu\nu}{F}^{\mu\nu}+j_\mu
A^\mu $. This Lagrangian is Lorentz invariant and is responsible
for the confinement of the global defects carrying electrical
current density $j_\mu$.} The effective Lagrangian ${\cal
L}^\phi_{eff}$ is the key point of investigation of {\it global
defects}. These global defect structures firstly found in
\cite{bmm} add to a list of other extended objects presenting
similar confinement profile as for instance Dp-branes in
string/M-theory \cite{witten97}, monopoles \cite{sbwit,kneipp1}
and parallel domain walls \cite{bbf101} in standard field theory.
And according to the Olive-Montonen conjecture \cite{olive}, one
may find models in which extended objects may present a dual role,
changing their contents with ordinary particles in the dual model:
the weak coupling phase gives rise to defect solutions, and
duality connects this phase with the strong coupling phase, which
exposes confinement of particles. An example of this is the
confinement of electric charge, which is connected to condensation
of magnetic monopoles \cite{sbwit}. The global defect structures
that we use \cite{bmm} are stable structures extended along the
radial dimension, and they appear in models described by a single
real scalar field. These defects do not require the introduction
of gauge fields, as it happens, for instance, to monopoles and
cosmic strings. And also, they do not violate the Derrick-Hobard
theorem \cite{H,D} (see also \cite{J}), because of the potential
that we consider in \cite{bmm}. Indeed, such defects require less
degrees of freedom of the effective theory than the global
monopole and the global cosmic string. Thus, these objects can
appear in a confined phase even though many relevant degrees of
freedom of the theory are frozen or simply dropped out due to a
sequence of spontaneous symmetry breaking. We organize our work as
follows. In the next Sec.~\ref{confD} we present the basic ideas,
and we study explicitly the confining behavior in $D\!=\!3$
dimensions. Also, in Sec.~\ref{emd} we work in $D\!=\!3$
dimensions, and we calculate the electric and magnetic energies of
the defects, and we show that for suitable choice of parameters,
the model may also support dyonic-like structures. In
Sec.~\ref{concu} we present our comments and conclusions.

\section{Confinement from global defects in $D$ dimensions}
\label{confD}

We consider a QCD-like effective Lagrangian which can be written
in terms of {\it Abelian} gauge fields provided we consider the
theory in a color dielectric medium, characterized by a color
dielectric function $G(\phi)$ with suitable asymptotic behavior. {
In a medium which accounts mainly for one-gluon exchange, the
gluon field equations linearize and are formally identical to the
Maxwell equations \cite{td,willets}. Through the color dielectric
function $G(\phi)$ the dynamics of $\phi$ is coupled to an
averaged gauge field $A_\mu^a$ which has only low-momentum
components \cite{rosina}. In addition, it is believed that in the
infrared limit there is an Abelian dominance \cite{chris}. In such
a limit one can fix the non-Abelian degrees of freedom in
non-Abelian $SU(N)$ theories in the {\it maximal Abelian gauge},
leaving a residual $U(1)^{N-1}$ gauge freedom \cite{thooft,weise}.
Results in QCD lattice has been shown that the Abelian part of the
string tension accounts for $92\%$ of the confinement part of
static lattice potential \cite{shiba,bali}. Thus, it suffices to
consider only the Abelian part of the non-Abelian strength field,
i.e., $F^a_{\mu\nu}\!=\!\partial_\mu A^a_\nu-\partial_\nu
A^a_\mu$. Furthermore, without loss of generality we can suppress
the color index ``$a$" if we take an Abelian external color
current density $j_a^\mu$
\cite{slusa,dick1,dick2,chabab1,chabab2}.} We account to these
facts and we write down the following low energy effective Abelian
Lagrangian \ben\label{QCD} {\cal
L}&=&-\frac{1}{4}G(\phi)F_{\mu\nu}F^{\mu\nu}-j_\mu A^\mu
+\frac{1}{2}\partial_\mu\phi\partial^\mu\phi
-V(\phi)\nonumber\\
&&+ \bar{\psi}(i\gamma^\mu\partial_\mu-m-f(\phi))\psi, \een Notice
that the spinor field $\psi$ is coupled to the scalar field via
the standard Yukawa coupling term $\bar{\psi}f(\phi)\psi$. The
first term in (\ref{QCD}) can also be regarded as a by-product of
Kaluza-Klein compactifications of supergravity theories, where
$G(\phi)$ in general is a function of both dilaton and moduli
fields \cite{mts}.

We now look for background solutions in the bosonic sector of
(\ref{QCD}). The equations of motion for static fields,
$j^\mu=(\rho,0,...,0)$ and $A_\mu=(A_0,...,0)$ in arbitrary $D$
dimensions are given by \ben\label{eqm1}
\nabla^2\phi=V_\phi-\frac{1}{2}|\overrightarrow{\nabla}U|^2G_\phi,
\\
\label{eqm2}
\overrightarrow{\nabla}.[G(\phi(\vec{r}))\overrightarrow{\nabla}U]=
-\rho(\vec{r}),
\een
where $U=A_0$ and the subscript $\phi$ on
any function means derivative with respect to $\phi$. It suffices
for now to turn on the electric field alone. As we shall see below,
the magnetic behavior can be found easily by using the results of
the electric case.

We now suppose that the fields engender
radial symmetry, i.e., $\rho=\rho(r)$, $\phi=\phi(r)$ and
$U=U(r)$ with $r=\sqrt{x_1^2+...+x_D^2}$. In this case
the equations (\ref{eqm1}) and (\ref{eqm2}) read
\ben\label{eqm12}
&&\frac{1}{r^{D-1}}\frac{d}{dr}\left(r^{D-1}\frac{d\phi}{dr}
\right)=V_\phi-\frac{1}{2}\left(\frac{dU}{dr}\right)^2G_\phi
\\
\label{eqm22}
&&\frac{1}{r^{D-1}}\frac{d}{dr}\left(r^{D-1}G(\phi({r}))\frac{dU}
{dr}\right)=-\rho(r). \een In order to find classical solutions of
the equations of motion we take advantage of the first-order
differential equations that appear in a way similar to
Bogomol'nyi's approach, although we are not dealing with
supersymmetry in this paper. That is, we follow
Ref.~{\cite{bbf101}} and we consider the potential $V$ as
\ben\label{pot}
V(\phi,r)=\frac{1}{r^{2D-2}}\left\{\frac{1}{2}W_\phi^2-\frac{Q_e^2}{2G}
\right\}, \een where we have defined the electrical charge of a
defect with a typical radius $R$ as
$Q_e=-\int_0^R\rho(r)r^{D-1}dr$. { Potentials breaking Lorentz
invariance as above, have been introduced in the literature, e.g.,
in the context of supergravity \cite{mts}, dynamics of embedded
kinks \cite{salemvash} and scalar field in backgrounds provided by
scalar fields \cite{olum}}.


 With this choice, the equations of motion can be
solved by the following first-order differential equations
\ben\label{eqf1} &&r^{D-1}\frac{d\phi}{dr}=\pm W_\phi
\\
&&\label{eqf2} r^{D-1}G\frac{dU}{dr}=Q_e. \een We notice that the
scalar field $\phi(r)$ is now decoupled from the gauge field
$U(r)$. { According to the equations (\ref{eqf1})-(\ref{eqf2}) we
see that the dynamics of the gauge field does not affect the
dynamics of the scalar field. On the other hand, the scalar field
develops a background field that affects the gauge field dynamics
through equation (\ref{eqf2}) --- recall $G\equiv G(\phi)$. Same
happens to the fermion field $\psi$ for which $\phi$ develops a
bosonic background field via equation (\ref{fermionEOM}) below.
Thus the stability of the new global defects investigated in
\cite{bmm} remains valid here. The topological charge for such
defects is given by \ben Q_T^D=\int
d\vec{r}\rho_D^2=(-1)^D(D-1)!\Omega_D\Delta W, \een where
$\Omega_D$ is the $D$-dependent angular factor and $\Delta W=\pm
W[{\phi(r\to\infty)}]\mp W[{\phi(r=0)}]$ as has been shown in
\cite{bmm}}.

The functions $W$, $\rho$ and $G$ are to be chosen properly in
order to describe confinement of defects.

The color charge density $\rho$ is determined by the fact that
defects get charged by fermions. In order to make this clear let
us analyze the localization of fermion zero modes on the defects
\cite{jack}. The variation of ${\delta{\cal
L}}/{\delta\bar{\psi}}=0$ gives the equation of motion for the
Dirac fermion. The fermion equation of motion in $D$-dimensional
spacetime, in spherical coordinates is \ben\label{fermionEOM}
i\gamma^0\partial_0\psi+i\gamma^r\partial_r\psi+
q\gamma^0A_0(r)\psi\!-\![m\!+\!f(\phi(r))]\psi\!=\!0 \een where
$\psi$ depends only on $t$ and the radial coordinate $r$. Also,
$\gamma^r$ is in general a linear combination of standard
Minkowski gamma matrices $\gamma^\mu$. However, we can regard
$\gamma^r$ here as a Minkowski gamma matrix itself provided we
transform $\psi$ under a local rotating frame \cite{villalba}. We
have chosen the Yukawa coupling as
$f(\phi(r))=-m+\widetilde{W}_{\phi\phi}(\phi(r))$, where $\phi(r)$
is the bosonic background solution of the scalar field. Moreover,
we follow \cite{bmm} and we choose
$\widetilde{W}_{\phi\phi}=r^{1-D}W_{\phi\phi}.$

We look for solutions of (\ref{fermionEOM}) by assuming
{\it a priori} that the fermion field is localized on the defect and
lives in a region of thickness $2R$. As long as $R$ goes to zero,
the color charge density tends to a delta function; the color
electric potential $A_0(r)$ becomes very strong with the behavior
of a Coulomb potential and turns out to be approximately constant,
i.e., $\partial_rA_0\approx0$. Alternatively, we could gauge away
the field $A_\mu$ in (\ref{fermionEOM}) using the ``pure gauge''
choice $A_\mu=\partial_\mu\Lambda$ and
$\psi(x_\mu)=\exp{[-iq\Lambda(x_\mu)]}h(x_\mu)$. We account for
this fact considering the ansatz
\ben\label{ansatz}
\psi=e^{iqA_0(r)t}h(r)\epsilon_\pm,
\een
where $\epsilon_\pm$ is a
constant spinor and $A_0(r)$ is an approximately constant color
electric potential in a region of radius $R$. We substitute
(\ref{ansatz}) into (\ref{fermionEOM}) and use the fact that
$\gamma^r\epsilon_\pm=\pm i\epsilon_\pm$ to find the zero mode
solution $h_0(r)=C\exp{[\mp\int^r
{r'}^{(1-D)}W_{\phi\phi}(r')dr']}$, where $C$ is a normalization
constant. We now use Eq.~(\ref{ansatz}) to write the spinor solution,
and so we get
\ben\label{fermSol}
\psi_0(r,t)=e^{iqA_0(r)t}[Ce^{\mp\int^r{r'}^{(1-D)}
W_{\phi\phi}(r')dr'}]\epsilon_\pm
\een
Using these general solutions, the localized charge on the global
defects due to the fermionic carriers is given by the current density
$j_\mu=q\bar{\psi}_0\gamma_\mu\psi_0$. The color charge
density is $\rho=j_0=q\psi_0^\dagger\psi_0$ and so, using
the general fermion solution (\ref{fermSol}) we get
\ben\label{rho}
\rho(r)=qC_0e^{\mp
2\int^r{r'}^{(1-D)}W_{\phi\phi}(r')dr'},
\een
where $C_0$ is a normalization constant.

The color dielectric function $G$ is chosen to have an
anti-screening property that acts against the separation of
defects. Such behavior can be summarized as follows
\ben\label{colord1} G(\phi(r))\to1, r\leq1, \qquad G(\phi(r))\to0,
r\gg1, \een as we have anticipated in the introduction.

As it was already shown in \cite{bmm}, for $D\geq2$ the global
defect solutions are obtained after mapping the $D$-dimensional
problem into an one-dimensional model. Such a mapping is obtained
by identifying the left-hand side of equation (\ref{eqf1}) with an
ordinary derivative \ben
r^{D-1}\frac{d\phi(r)}{dr}=\frac{d\phi(x)}{dx} \een where $x$ is a
non-compact coordinate. It is not hard to see that this implies
into a map between $r$ and $x$ of the form $dx=\pm r^{1-D}dr$. In
the case of $D=2$ one has $x=\ln{r},$ in a way such that one maps
$r\in [0,\infty)$ to $x\in (-\infty,\infty)$. This is an example
in which the whole coordinate $r$ is mapped to the whole
coordinate $x$. However, this is not the case for $D\geq3$, since
the map is now $x=\pm[1/(2-D)](1/r^{D-2})$. This maps $r\in
[0,\infty)$ to either $x\in (-\infty,0]$ (upper sign) or to $x\in
[0,\infty)$ (lower sign), i.e., the coordinate $x$ is now
``folded'' into a half-line ${\mathbb{R}}/{\mathbb{Z}_2}$. As we
shall see below, this imposes new conditions on the way one
chooses the function $W$. In the two-dimensional case we have
chosen $W$ such that we have found defect solution on the
coordinate $x$ connecting vacua at $(-\infty,\infty)$. For
$D\geq3$, since we have a boundary at $x=0$ the defect should
connect vacua at ($-\infty,0$] or [$0,\infty$). In order to
fulfill this requirement the function $W$ should have a critical
point at zero. Thus, we follow \cite{bmm} and we introduce another
$W$, in the form \ben\label{Wp}
W_\phi=(\phi^{\frac{p-1}{p}}-\phi^{\frac{p+1}{p}}). \een Here $p$
is integer, $p=1,2,...$ For $p>1$ one gets $W_\phi(0)=0$, which
shows that $\phi=0$ is a critical point of $W$. Extensions for two
scalar fields with $p\!=\!1$  are investigated in
\cite{bb,bbb,bb2000}. Substituting this $W$ into Eq.~(\ref{eqf1})
(considering the lower sign) one finds the following defect
solution \ben\label{solDp}
\phi(r)=\tanh^p{\left[\frac{1}{p}\left(\frac{r^{2-D}}{D-2}\right)\right]}.
\een For $D\geq3$ and $p=4,6,...$ this solution connects the
vacuum $\phi=1$ to the vacuum $\phi=0$ as $r$ goes from 0 to
$\infty$.

\subsection{Confinement in $D=3$ spatial dimensions}

\begin{figure}[ht]

\includegraphics[{height=5.0cm,width=7.0cm}]{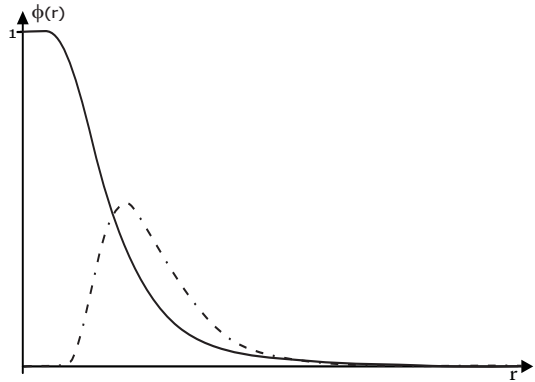}
\caption{The radial defect $\phi(r)$ and the color charge density
$\rho(r)$ (the dot-dashed line) for $D=3$ and $p=4$. The leading
term of $\rho(r)$ in the asymptotic limit is $\sim1/r^6$. The
color charge is localized around the radius $R$ of the defect
which approaches to its core for $p$ very large.}\label{fig2}
\end{figure}

We now turn attention to specific models, where we power the
scalar matter contents to give rise to confinement in our
four-dimensional universe, which means to consider only $D\!=\!3$
spatial dimensions. The solution (\ref{solDp}) turns out to be
\ben\label{sol23} \phi(r)=\tanh^p{(1/p\,r)}, \een Let us discuss
about the color charge density associated to this defect. We
substitute Eq.~(\ref{sol23}) into Eq.~(\ref{rho}), and we perform
the integral on the exponential analytically, for the upper sign,
to get \ben \label{rho3D} \rho(r)=q\,c(p)[\tanh{(1/p
r)}]^{2p-2}{\rm sech}^4{(1/p r)}, \een where $c(p)$ is a
normalization constant which increases monotonically with
$p=4,6,...$. The behavior of this color charge density for $p=4$
is depicted in Fig.~\ref{fig2}. Notice that $\rho(r)$ falls-off to
zero as $r$ goes to infinity. The leading term of $\rho(r)$ given
in (\ref{rho3D}) for very large $r$ (or very large $p$) is
$(1/r\,p)^{2p-2}$ or $1/(4r)^6$ for $p=4.$ This ensures that the
fermion solution (\ref{fermSol}), with the minus sign in the
exponential, is normalizable and localized on the defect
(\ref{sol23}). Thus, there is a localized charge distribution on
the defect. This charge density in $D\!=\!3$ in the limit of $p$
very large is given by \ben\label{charge}
\rho(r)=q\,c(p)\left(\frac{1}{r\,p}\right)^{2p-2}, \een which
turns out to be a delta function in the limit $p\to\infty$ through
even values. The distribution (\ref{charge}) is a good
approximation of (\ref{rho3D}) as long as the radius $R$ of the
defect, where $R=(1/p)\,\mbox{arctanh}[(1/2)^{1/p}]$, is very
small. Of course, this happens for $p$ even and very large. In
order to make calculations analytical, below we will use
Eq.~(\ref{charge}) to represent the charge density.

A color dielectric function $G$ for this case with the suitable
behavior displayed in Eq.~(\ref{colord1}) can be given by
\ben\label{G2}
G(\phi)=\left[\frac{\phi^2}{\phi^2+(A_\alpha)^l(\phi^{\beta/p}+
1)/2}\right]^{2\alpha},
\een
where $l,\,\alpha>0$, $\beta<0$ and $A_\alpha=4\alpha-1$.
Now, we substitute Eq.~(\ref{sol23}) in the limit of large $p,$
and we use Eqs.~(\ref{charge}) and (\ref{G2}) into (\ref{eqf2}) to
find
\ben\label{eqr3.3}
&&r^2\left\{\frac{1}{1+(A_\alpha)^l[(rp)^{-\beta+2p}+(rp)^{2p}]/2}
\right\}^{2\alpha}\frac{dU}{dr}\nonumber
\\
&&=\frac{q}{2}\frac{c(p)}{p}\left(\frac{1}{rp}\right)^{2p}. \een
In the case $2\alpha=1$ we take the leading term in $p$ to get
\ben \frac{dU}{dr}=\frac{q}{4}\frac{c(p)}{p}\left(\frac{1}{r^2}+
p^{-\beta}r^{-\beta-2}\right). \een This is the color electric
field in $D=3$ dimensions. The color electric potential obtained
by integrating over the radial coordinate $r$ is \ben\label{pot2}
U(r)=\frac{q}{4}\frac{c(p)}{p}\left[-\frac{1}{r}+f_\beta(r)\right].
\een The first term is the well-known Coulomb potential and the
second term comprises the confining part of the potential. This
part $f_\beta(r)$ can be given in several distinct forms: for
$\beta=-1$ we obtain \ben f_\beta(r)=p\ln{r}. \een On the other
hand, for $\beta\neq-1$, we get the following formula \ben
f_\beta(r)=-p^{-\beta}\,\frac{r^{-\beta-1}}{\beta+1}. \een In this
case we have, for $\beta=-3/2$, $f_\beta(r)\sim\sqrt{r}$, and this
reproduces the confining part of the Motyka-Zalewski potential
\cite{MZ,Z}; for $\beta=-2$, $f_\beta(r)\sim{r}$, and now we get
the well-known linear confining part of the Cornell potential; and
for $\beta=-3$, $f_\beta(r)\sim{r^2}$. For $\beta>-1$ the
potential does not confine. As we can see, in the linear regime
the tension of the {\it QCD string} \cite{td,willets} is very
strong and scales as $\sim p^2$.

\section{Electric, magnetic and dyonic defects}
\label{emd}

The electrically charged defects that we have found can be used to
give rise to other solutions. To do this, let us now  investigate
the total energy of the global defect found in $D\!=\!3$
dimensions. It can be cast to the form \cite{mts}
\ben\label{emass} E_e=\int_{1/p}^\infty
dr\;r^2\left[\frac{1}{2}\left(\frac{d\phi}{dr}\right)^2+
\frac{Q_e^2}{2\,r^4\,G}+V \right], \een where we integrate from
$1/p,$ for $p$ very large, to conform with the approximation we
have done for the charge density [see Eq.~(\ref{charge})], and
with the size of the defect. In the above expression, the second
term is the color electric contribution, $Q_e$ is the color
electric charge of the defect and $G$ is the color dielectric
function. The scalar potential with $W_\phi$ chosen as in
Eq.~(\ref{Wp}), goes to zero as $p$ (even) becomes very large. In
this limit our model approaches the model studied recently in
Ref.~{\cite{slusa2}}. In the limit of large $p$ the solution
(\ref{sol23}) and $G$ given as in Eq. (\ref{eqr3.3}) turn out to
be \ben\label{phiG} \phi(r)&\simeq&\left(\frac{1}{r\,p}\right)^p,
\\
\label{phiG2}
G(\phi(r))&\simeq&\left(\frac{1}{1+(A_\alpha)^l\,r^{2p}\,
p^{2p}}\right)^{2\alpha}, \een where $A_\alpha=4\alpha-1$ and $l$
are positive parameters of the approximated color dielectric
function. Now, we substitute Eqs~(\ref{phiG}) and (\ref{phiG2})
into Eq.~(\ref{emass}) to get to the energy \ben\label{emass2}
E_e=\frac{1}{2}\frac{(A_\alpha)^{l/2p}\,p\,Q_e^2\,\Gamma{\left(-2\,\alpha+
\frac{1}{2p}\right)}\,\Gamma{\left(-\frac{1}{2p}\right)}}{2p\,
\Gamma{(-2\,\alpha)}}. \een Since the parameter $p$ is assumed to
be very large, the formula above provides a simple expression for
the energy which is given by
$E_e=-\frac{1}{2}(A_\alpha)^{l/2p}\,p\, Q_e^2$. Notice that the
energy is very large because $p$ is very large.

The above result can be extended to magnetic solutions.
Differently of the lines followed by Ref.~{\cite{slusa2}}, here we
do not consider the Wu-Yang SU(2) magnetic monopole solution of
the non-Abelian sector of the theory; instead, we keep dealing
with the Abelian sector and we assume that the global defect that
we have found in $D\!=\!3$ is charged by a {\it total magnetic
charge} $Q_m$. We make use of the results for the electric case to
get the magnetic energy \ben \label{hmass2}
E_m=\frac{1}{2}\frac{(A_\alpha)^{l/2p}\,p\,Q_m^2\,\Gamma{\left(2\,\alpha+
\frac{1}{2p}\right)}\,\Gamma{\left(-\frac{1}{2p}\right)}}{2p\,
\Gamma{(2\,\alpha)}}. \een The magnetic energy is easily found by
making the transformation $Q_e\to Q_m$ and $G\to1/G$ into
Eq.~(\ref{emass}) --- see \cite{mts} for details. As in the
electric case, for large $p$, the magnetic energy tends to be of
the form $E_m=-\frac{1}{2}(A_\alpha)^{l/2p}\,p\, Q_m^2$. Notice
that we are changing the dielectric function to a ``magnetic
permeability.'' According to (\ref{phiG2}), we see that this
transformation is equivalent to the change $\alpha\to-\alpha$. We
can refer to the two charged defects as ``electric defect'' and
``magnetic defect'', respectively.

The above electric and magnetic structures may be used to ask for
the presence of ``dyonic'' configurations. As one knows, a dyon is
an object with both electric and magnetic charges. Thus, let us
consider the possibility of a dyonic structure with mass
\ben\label{dyon} M_D=|E_e+E_m|\leq|E_e|+|E_m|=M_e+M_m. \een The
stability of the dyonic object would be ensured by the inequality
$M_D<M_e+M_m$. In supersymmetric theories the dyon is a BPS state
whose mass is related to the central charge of the supersymmetric
algebra. The stability of such BPS state depends crucially on the
value of its parameters \cite{sbwit}. Thus, in our case we can
think of a similar way of getting stability of the dyon state with
mass $M_D$. To do this, we see that for a suitable value of the
parameter $l$ the energy of the magnetic defect (\ref{hmass2})
becomes imaginary. In this way, the sum $E_e+E_m$ produce a vector
in the complex plane and then $M_D$ is the length of this vector
whose magnitude is lesser than the sum of its components.
Alternatively, for $l/2p$ being a semi-integer, and for
$\alpha\to-\alpha$, $(A_\alpha)^{l/2p}\to
i(\tilde{A}_\alpha)^{l/2p},$ (with $\tilde{A}_\alpha=|4\alpha+1|$)
we get $E_m\to i\tilde{E}_m$. In this case $M_D$ is given by the
formula \ben \label{dyon2}
M_D=\sqrt{E_e^2+\tilde{E}_m^2}=\frac{1}{2}p\sqrt{q_e^2+q_m^2},
\een where the charges $q_{e}$ and $q_{m}$ are defined in terms of
$E_e$ and $E_m$ in the limit of very large $p$ as \ben
q_e=(A_\alpha)^{l/2p}\, Q_e^2,
\\
q_m=(\tilde{A}_\alpha)^{l/2p}\, Q_m^2, \een We notice that
$q_{e}\leftrightarrow q_{m}$ as long as
$\alpha\leftrightarrow-\alpha$ and $G\leftrightarrow 1/G$.

\section{Comments and Conclusions}
\label{concu}

We summarize this work recalling that we have studied how the presence
of new global defect structures could act to confine in an simplified,
Abelian model where the scalar field self-interacts nontrivially, responding
for changing the dielectric properties of the medium. To make the model more
realistic, we have added fermions, which interact with the scalar field
through an Yukawa coupling, similar to couplings required in a
supersymmetric environment.

We have investigated a model in $D\!=\!3$ spatial dimensions that
seems to respond to confinement very appropriately. Similar
investigations can perfectly be addressed in any $D$ spatial
dimensions. In particular, in $D\!=\!1$ we could use the
superpotential (\ref{Wp}) for $p\!=\!3,5,...,$ which was shown to
give rise to defect solutions of the form of 2-kink solutions
\cite{bmm}, which the kinks separated by a distance which
increases with increasing $p\!=\!3,5,...$ This feature has been
further explored in Ref.~{\cite{bfg}}, where one couples the model
to gravity in $(4,1)$ spacetime dimensions in warped spacetime
with one extra dimension. There one also finds results which
highlight the fact that the superpotential (\ref{Wp}) gives rise
to 2-kink defect structures for $p\!=\!3,5,...$ These solutions
could be used in the present context, to investigate confinement
with these 2-kink defects. Investigations on this appear
interesting because it could offer an alternative to
Ref.~{\cite{bbf101}}, using a simpler model which requires a
single real scalar field to generate 2-kink structures.

In $D\!=\!3$ dimensions, we have shown that the monopole-like,
electrically charged global defect can be used to generate
magnetically charged global defect. This is implemented by
essentially changing the ``color'' dielectric function $G(\phi)$
to its inverse $1/G(\phi).$ This feature has inspired us to search
for dyons, for dyonic-like structures which should be stable bound
states of electrically and magnetically charged global defects,
whose mass should be given in terms of the electric and magnetic
charges.

\acknowledgments

We would like to thank Adalto Gomes, Maciek Slusarczyk and Andrzej
Wereszczynski for useful discussions, and CAPES, CNPq, PROCAD and PRONEX
for partial support.  WF also thanks FUNCAP for a fellowship.


\begin{thebibliography}{99}
\bibitem{footnote}{A similar
procedure to construct an effective Lagrangian by substituting
solutions of equation of motion into the original Lagrangian has
been considered, e.g., in Ref.~\cite{olum}.}
\end{thebibliography}
\end{document}